\documentclass{JHEP3}
\usepackage{amsmath}
\usepackage{cite}
\usepackage{epsfig}
\usepackage{xspace}
\usepackage{graphics}
\usepackage[latin1]{inputenc}

\def\hide#1{}

\newcommand{\pythia}{P\scalebox{0.8}{YTHIA}\xspace}

\newcommand{\as}{\ensuremath{\alpha_{\mathrm{s}}}}

\def\mrm#1{\mathrm{#1}}

\def\sub#1{\ensuremath{_{\mrm{#1}}}}

\def\f2d3{\ensuremath{F_2^{\mrm{D}3}}}

\def\done#1{}

\providecommand{\eqref}[1]{eq.~(\ref{#1})\xspace}
\renewcommand{\eqref}[1]{eq.~(\ref{#1})\xspace}
\newcommand{\eqsref}[1]{eqs.~(\ref{#1})\xspace}

\newcounter{aenumct}

\newcounter{ienumct}

{\end{list}}

\newcounter{enumct}

\usepackage{axodraw}
\usepackage{epsfig}
\usepackage{amssymb}
\usepackage{mathrsfs}
\usepackage{mathcomp}
\usepackage{cite}
\usepackage[latin1]{inputenc}

\def\pmb#1{{\mbox{\boldmath$#1$}}}
\def\ie{\emph{i.e.}}
\def\eg{\emph{e.g.}}
\def\vs{\emph{vs.}}

\keywords{Small-$x$ physics, Saturation, Diffraction, Dipole Model, DIS}
\preprint{LU-TP 07-25\\
}

\title{Diffractive Excitation in DIS and $pp$ Collisions}

\author{Emil Avsar, Gösta Gustafson and Leif Lönnblad\\
  Dept.~of Theoretical Physics,
  Sölvegatan 14A, S-223 62  Lund, Sweden\\
  E-mail: \email{emil@thep.lu.se}, \email{gosta@thep.lu.se}
    and \email{leif@thep.lu.se}}

  \abstract{We have in earlier papers presented an extension of
    Mueller's dipole cascade model, which includes subleading effects
    from energy conservation and running coupling as well as colour
    suppressed effects from pomeron loops via a ``dipole swing''. The
    model was applied to describe the total cross sections in $pp$ and
    $\gamma^*p$ collisions.

    In this paper we present a number of improvements of the model, in
    particular related to the confinement mechanism.  A consistent
    treatment of dipole evolution and dipole--dipole interactions is
    achieved by replacing the infinite range Coulomb potential by a
    screened potential, which further improves the frame-independence
    of the model.

    We then apply the model to elastic scattering and diffractive
    excitation, where we specifically study the effects of different
    sources for fluctuations. In our formalism we can take into
    account contributions from all different sources, from the dipole
    cascade evolution, the dipole--dipole scattering, from the
    impact-parameter dependence, and from the initial photon and proton
    wavefunctions. Good agreement is obtained with data from the
    Tevatron and from HERA, and we also present some predictions for
    the LHC.}

\begin{document}

\sloppy

\section{Introduction}
\label{sec:intro}

In high energy $pp$ scattering the cross section for parton--parton
subcollisions becomes larger than the total cross section. This means
that on average there are more than one subcollision in a single
event, and it was early suggested that hard subcollisions dominate the
features of high energy scattering and are the cause of the rising
cross section.  This is also the basic assumption in the model by
Sj\"ostrand and van Zijl \cite{Sjostrand:1987su} implemented in the
\pythia event generator, which is able to describe many features of
high energy collisions.

That perturbative dynamics dominate high energy collisions is also
supported by the large intercept of the BFKL pomeron. Via unitarity
and the AGK cutting rules, the large subcollision cross section and
high probability for multiple collisions have also strong implications
for diffraction. Hard diffraction was first observed by the UA8
collaboration at the CERN Sp\={p}S collider \cite{Bonino:1988ae}, and
has later been studied in much more detail both at the Tevatron (see
\eg\ \cite{Abe:1997rg,Abbott:1999km}) and at HERA (see
\eg\ \cite{Derrick:1993xh,Ahmed:1994nw}).

Multiple collisions, unitarity and saturation are conveniently studied
in terms of dipoles in transverse coordinate space. The dipole model by
Golec-Biernat and W\"usthoff (GBW)
\cite{Golec-Biernat:1998js,Golec-Biernat:1999qd} has successfully
described both $F_2$ and diffraction in DIS. Mueller's dipole cascade
model \cite{Mueller:1993rr,Mueller:1994jq,Mueller:1994gb} reproduces
the leading log (linear) BFKL equation, and includes also multiple
collisions and satisfies the unitarity constraint. The multiple
collisions correspond to pomeron loops.  Mueller's model includes,
however, only such loops which are cut in the particular Lorentz frame
used in the calculation, but not loops which are fully contained in
one of the individual dipole cascades. Many attempts (see for example
\cite{Iancu:2005dx, Kozlov:2006cg} and references therein) have been
presented including \eg\ $2\to 1$, $2\to 4$ or more
complicated dipole vertices, but so far no explicitly
frame-independent formalism has been presented.

As discussed in some detail below, an important part of the NLL 
corrections to the BFKL equation are
related to energy conservation. In a series of
papers \cite{Avsar:2005iz, Avsar:2006jy} we have developed Mueller's
model to include both effects of energy--momentum conservation and
effects of pomeron loops and saturation inside the cascade evolution
via a $2\rightarrow 2$ dipole transition, called a dipole swing. The
swing does not reduce the number of dipoles, rather the saturation
effect is achieved as the "new" dipoles are smaller, and therefore
have smaller cross sections.  Although not explicitly frame
independent, the numerical result is almost independent of the frame
used for the calculations. With a simple model for the proton in terms
of three dipoles, the Monte Carlo implementation also reproduces the
total cross section both for DIS at HERA and for $pp$ scattering from
ISR energies to the Tevatron.

In this paper we will first make some technical improvements related
to confinement, and then use the model to study diffractive scattering
at HERA and the Tevatron. The perturbative calculation has some
problems in the IR region, especially with a running coupling, and a
cutoff for large dipoles is essential for the frame independence and
the good agreement with data. We here propose to treat this effect of
confinement by everywhere replacing the Coulomb colour-electric
potential by a screened Yukawa potential.

Both the screening length and the size of the initial proton wave
function are determined by the confinement mechanism.  In the MC
implementation these two quantities are assumed to be of the same
size.  This implies that for $pp$-collisions and DIS at high $Q^2$ the
model has only two tunable parameters; besides $\Lambda_{QCD}$ only
the confinement scale denoted by $r_{max}$. For smaller $Q^2$ the
result is, however, also sensitive to the quark masses in the virtual
photon wave function. The value of $r_{max}$ turns out to be very
important in order to obtain the correct normalization for
$\sigma_{tot}$ in $pp$ collisions, but the increase of $\sigma_{tot}$
with the center of mass energy is found to be much less sensitive to
this parameter. Once $r_{\max}$ is fixed to obtain the correct
normalization for $\sigma_{tot}$ in $pp$ scattering, the DIS cross
section is obtained without any further changes.

Our treatment of elastic scattering and diffractive excitation is
based on the eikonal approximation and the Good and Walker picture
\cite{Good:1960ba}. The result is determined by the fluctuations in
the collision process originating from the initial wave functions of the
proton and the virtual photon, from the dipole cascades and from the
dipole--dipole scattering probability.  In our formalism all these
different components give important contributions.  One result of this
is that the impact parameter profile is less steeply falling, \ie\
less black and white and more ``grey'', than in models where the
dominant fluctuations are assumed to come from fluctuations in the
impact parameter, $b$.

The distribution in the mass, $M_X$, of the diffractive state can be
obtained by a study of the collision in different Lorentz frames, as
discussed by Hatta et al.\ \cite{Hatta:2006hs}. (It is here essential
that we have a frame-independent formalism.) However, in addition to
the fluctuations included in this reference and in the GBW approach,
we also include fluctuations in the evolution of the proton 
target.

In section \ref{sec:dipmod} we review briefly the dipole cascade
model, discuss the modification of the confinement effect, and
demonstrate the frame independence of the model. In section
\ref{sec:diff} we discuss the formalism for elastic and diffractive
scattering, and the effects of the different sources of fluctuations
in the collision process.  Our results are presented in section
\ref{sec:res}, and the conclusions in section \ref{sec:conc}.

\section{Dipole Model and Frame Independence}
\label{sec:dipmod}

We will in this section briefly discuss the cascade model presented in
refs.~\cite{Avsar:2005iz, Avsar:2006jy}, describe the modified
treatment of confinement, and demonstrate the frame independence by
showing some quantitative examples.

\subsection{Mueller's cascade model}

The model is based on Mueller's dipole formalism
\cite{Mueller:1993rr,Mueller:1994jq,Mueller:1994gb} in which the
small-$x$ evolution is interpreted in terms of a dipole cascade.  The
probability per unit rapidity $Y$ that a dipole $(\pmb{x},\pmb{y})$
emits a gluon at transverse position $\pmb{z}$ is here given by
\begin{eqnarray}
\frac{d\mathcal{P}}{dY}=\frac{\bar{\alpha}}{2\pi}d^2\pmb{z}
\frac{(\pmb{x}-\pmb{y})^2}{(\pmb{x}-\pmb{z})^2 (\pmb{z}-\pmb{y})^2},
\label{eq:dipkernel1}
\end{eqnarray}
and the evolution of the cascade agrees with the leading order BFKL
evolution. As a consequence, the total number of dipoles grows
exponentially.  This also implies a strong growth for the total cross
section which, however, is tamed by taking multiple dipole
interactions into account. The scattering probability between two
elementary colour dipoles with coordinates $(\pmb{x}_i,\pmb{y}_i)$ and
$(\pmb{x}_j,\pmb{y}_j)$ respectively, is given by
\begin{equation}
  f_{ij} = f(\pmb{x}_i,\pmb{y}_i|\pmb{x}_j,\pmb{y}_j) =
  \frac{\as^2}{8}\biggl[\log\biggl(\frac{(\pmb{x}_i-\pmb{y}_j)^2
    (\pmb{y}_i-\pmb{x}_j)^2}
  {(\pmb{x}_i-\pmb{x}_j)^2(\pmb{y}_i-\pmb{y}_j)^2}\biggr)\biggr]^2.
\label{eq:dipamp}
\end{equation}
Since Mueller's model is formulated in transverse coordinate space,
multiple scatterings can be included in an eikonal approximation, and
a unitarised expression for the total scattering amplitude can be
obtained as
\begin{eqnarray}
T(\pmb{b}) = 1 - \mathrm{exp}\biggl ( - \sum_{ij} f_{ij} \biggr ).
\label{eq:T}
\end{eqnarray}

\subsection{Energy conservation}

The fast growth in leading order BFKL is much reduced by NLL effects.
As discussed in ref.~\cite{Salam:1999cn}, besides effects from the running 
coupling (which are also included in our simulations) the NLL corrections to 
BFKL evolution are related to the non-singular terms in the $g\rightarrow gg$
splitting function, and to the so called energy scale terms, 
and both these effects are related to the conservation of energy and
momentum.  It should be noted, however, that although these NLL
corrections to the evolution kernel are essential for the conservation
of energy and momentum, they are not sufficient to guarantee global
energy-momentum conservation, which is most easily achieved in a
computer simulation \cite{Andersen:2006sp}.

The LL BFKL result is obtained 
from the singular term $\sim 1/z$ in the gluon splitting function, which
dominates for small $z$.
The recoil for the gluon taking the fraction $(1-z)$ is here
neglected. However, for large $z$-values energy conservation 
implies that this recoil cannot be neglected.
(More precisely it is conservation of the positive light-cone 
component, as in a boost
invariant formalism the variables $x_{Bj}$ and $z$ are defined as fractions 
of $k_+ = k_0 + k_\mathrm{L}$.) If this suppression can be described by an
effective cutoff $z<a$ (with $a=\mathcal{O}(0.5)$), an integral over $z$ is
replaced as \cite{Andersson:1995ju}
\begin{equation}
\int_\epsilon^1 \frac{d z}{z} = \ln 1/\epsilon \,\rightarrow  
\int_\epsilon^a \frac{d z}{z} = \ln 1/\epsilon - \ln 1/a.
\end{equation}
As discussed in ref.~\cite{Gustafson:2001iz} the effect of the
non-singular terms in the splitting function is just such a suppression
of large $z$-values, with $\ln 1/a = 11/12$.

The so called energy scale terms are needed in order for a 
non-$k_\perp$-ordered ladder to look the same if evolved from the bottom up 
or from the top down. Thus a step down in $k_\perp$ must look exactly like a 
step up in $k_\perp$, if generated in the opposite direction. This implies that
the cascade must be ordered not only in $k_+$ (via the constraint $z<1$) but
also in the negative light-cone momentum $k_-$. This constraint is therefore 
also related to energy-momentum conservation, and it corresponds to what 
is commonly called the consistency constraint. As discussed in
refs.~\cite{Andersson:1995ju,Andersson:1998bx}, it has the consequence that
the pole 
$1/(1-\gamma)$ in the characteristic function $\chi(\gamma)$ is replaced by
$1/(1+\omega-\gamma)$. The minimum of $\chi(\gamma)$ 
is then shifted to $\gamma=(1+\omega)/2$, and thus a distance
$\omega/2$ further away from the poles. Expanding this effect to lowest order 
in $\bar{\alpha}$ gives the energy scale terms presented in 
ref.~\cite{Salam:1999cn}. 

Using the similarities between
Mueller's cascade model and the Linked Dipole Chain (LDC) model
\cite{Andersson:1995ju,Andersson:1998bx}, we implemented
energy--momentum conservation in the dipole cascade in
ref.~\cite{Avsar:2005iz}.  The $p_\perp$ of the partons was here
associated with the dipole sizes in coordinate space. Although the
number of dipoles still increases exponentially, the growth is
significantly reduced and the onset of saturation is delayed.

\subsection{Initial proton and photon wave functions}
\label{sec:wavefunctions}

\subsubsection{Photon}
\label{sec:photon}

The splitting $\gamma^* \to q\bar{q}$ can be calculated
perturbatively, and we use the well known leading order results for
longitudinally and transversely polarized photons:
\begin{eqnarray}
  \vert \psi_L(z,r,Q^2)\vert^2&=\frac{6\alpha_{em}}{\pi^2}\sum_f 
  e_f^2&\!Q^2z^2(1-z)^2K_0^2\left(r\sqrt{z(1-z)Q^2+m_f^2}\right) \nonumber \\
  \vert \psi_T(z,r,Q^2)\vert^2&=\frac{3\alpha_{em}}{2\pi^2}
  \sum_fe_f^2&\!\left\{[z^2+(1-z)^2](z(1-z)Q^2+m_f^2)K_1^2
    \left(r\sqrt{z(1-z)Q^2+m_f^2}\right)\right. \nonumber \\
  &&\quad+m_f^2\left.K_0^2\left(r\sqrt{z(1-z)Q^2+m_f^2}\right) \right\}.
  \label{eq:psigamma}
\end{eqnarray}
Here $z$ is the negative light-cone momentum fraction of the photon
carried by $q$, and $r$ is the transverse separation between $q$ and
$\bar{q}$, and we consider four active quark flavours, with an
effective light quark mass of 60 MeV and a charm mass of 1.4~GeV, as
described in \cite{Avsar:2007ht}.

Although the wave functions in \eqref{eq:psigamma} are well known,
there are still ambiguities in the initial phase of the evolution of
the dipole cascade. This problem is an unavoidable consequence of the
difficulty to reconcile the fundamentally quantum mechanical process
with the semiclassical approximation represented by the cascade
evolution.
 
For a given $W$ in DIS, the total rapidity range available for final
state particles is given by $Y=\ln W^2/m_0^2$, where $m_0$ is of the
order of\footnote{In the following the scale $m_0=1$~GeV will be
  omitted in writing logarithms.} 1~GeV. We use a Lorentz frame such
that an interval $Y_0$ is on the photon side and the remaining
interval $Y-Y_0$ is on the proton side. The kinematics is illustrated
in figure \ref{fig:triangle}, which shows the phase-space diagram for
a DIS event in the $(\ln p_\perp,y)$-plane. Here the positions of the
$q$ and $\bar{q}$ are also indicated.  Their distance in rapidity from
the photon end are given by $\ln p_\perp/z$ and $\ln p_\perp/(1-z)$
respectively, where $p_\perp$ is the transverse momentum of the quark
and the antiquark.

In the simulations we have assumed that gluon emission is possible
only at rapidities larger than the rapidity of both $q$ and $\bar{q}$.
We also identify $p_\perp$ with its typical value $2/r$. As can be
seen from figure \ref{fig:triangle}, this implies that the interval
allowed for the photon-initiated cascade is given by
\begin{equation}
Y_\gamma^{eff} = Y_0 - \mathrm{ln}\biggl ( \frac{p_\perp}{\mathrm{min}(z,1-z)} \biggr )
=Y_0 - \mathrm{ln}\biggl ( \frac{2}{r \mathrm{min}(z,1-z)} \biggr ).
\label{eq:ygammaeff}
\end{equation}

Thus $Y_\gamma^{eff}$ depends on both variables $r$ and $z$. The total
range for the evolution
\begin{equation}
Y_p + Y_\gamma^{eff} = \ln W^2 - \mathrm{ln}\biggl ( \frac{2}{r \mathrm{min}(z,1-z)} \biggr )
\label{eq:ygammaeffyp}
\end{equation}
may be larger or smaller than $\ln 1/x \approx \ln W^2 - \ln Q^2$. One
could also imagine other choices, but we want to emphasize that the
difference is subleading in $\ln W^2$, and therefore the optimal
choice cannot be determined from QCD with present techniques.

 \FIGURE{\includegraphics[angle=0, scale=0.7]{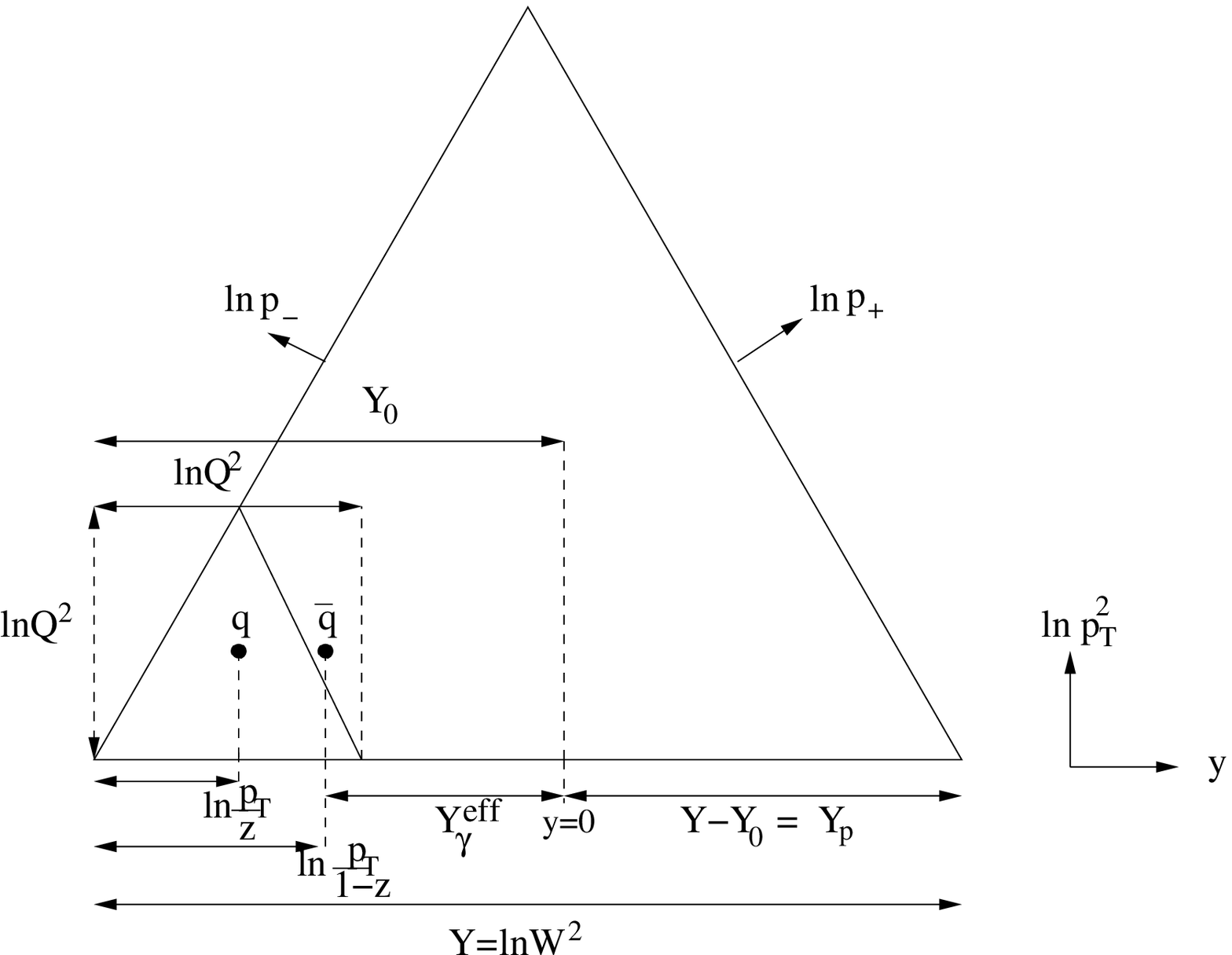}
   \caption{\label{fig:triangle} The available phase space for gluon
     emission in DIS.  We have here for simplicity assumed that $z >
     1-z$. The vertical dashed line labeled $y=0$ denotes the Lorentz
     frame in which the collision is studied. } }


Anticipating the discussion in the following sections, we want in
connection to figure \ref{fig:triangle} mention that a calculation of
the diffractive cross section corresponds to events which have a gap
at $y=0$ in this particular Lorentz frame. This means that the
diffractively excited photon is confined within the rapidity range
$Y_0$ corresponding to a mass limited by $M_{X,max}^2 \approx
\exp{Y_0}$.

\subsubsection{Proton}
\label{sec:proton}

The initial proton wavefunction can obviously not be determined by
perturbation theory, but has to be specified by a model. Our
assumption is that the initial proton can be described by three
dipoles in a triangular configuration, where the corners could be
associated with the three valence quarks. We admit that this model has
deficiencies, and it should not be used to study the particle
distribution in the proton fragmentation regions. The corner
connecting two dipoles in a triangle corresponds to an octet colour
charge, and not a quark triplet, and an alternative model could be
three dipoles connected by a junction in a Y-shaped configuration.
This would imply the need for extra assumptions about the behaviour of
the colour-neutral junction, and we also believe that the enhanced
radiation from the three colour octets might mimic radiation from
gluon contributions present in the proton structure also at low
virtuality.

We have tried two different triangular shapes. In the first one the
triangle is assumed to be equilateral. In this case the size of the
dipoles is assumed to be distributed according to a Gaussian
\begin{equation}
dP_1(\pmb{r}) = \mathcal{N}_1\, d^2\pmb{r}\, \mathrm{exp}\biggl
(-\frac{r^2}{r_{max}^2} \biggr ).
\label{eq:pwave1}
\end{equation}
We found in ref.~\cite{Avsar:2006jy} that this simple model gives a
very good agreement with data on cross sections for both $pp$
scattering and DIS.  To test the sensitivity to this simple
assumption, we have now also studied a model where the triangle is not
equilateral, but has an arbitrary shape. Also here the distribution is
given by a Gaussian in the sizes of the three sides, $r_1$, $r_2$ and
$r_3$,
\begin{equation}
dP_2(\pmb{r}_1,\pmb{r}_2,\pmb{r}_3) = \mathcal{N}_2\,d^2\pmb{r}_1d^2\pmb{r}_2d^2\pmb{r}_3
\, \mathrm{exp}\biggl (-\frac{r_1^2+r_2^2+r_3^2}{r_{max}^2} \biggr ) 
\delta(\pmb{r}_1+\pmb{r}_2+\pmb{r}_3).
\label{eq:pwave2}
\end{equation}
The results of the two models are very similar, and we will therefore
in the following mainly present results obtained with the simple model
given by $dP_1(\pmb{r})$.

The parameter $r_{max}$, which determines the initial dipole size in
the proton, is here assumed to be the same as the confinement scale in
the cascade evolution and the dipole--dipole scattering (see section
\ref{sec:conf} below). Along with $\Lambda_{QCD}$, it is one of the
essentially two free parameters of our model. We note, however, that
the variation of these two parameters have similar effects. Thus an increased
value for $r_{max}$ can be compensated by a reduced value for 
$\Lambda_{QCD}$ (and \emph{vice versa}), leaving the cross sections unchanged
as seen in figure \ref{fig:pptotrmaxlqcd}. We note in particular that
the energy dependence is rather insensitive to the parameters chosen.

\FIGURE[t]{
  \epsfig{file=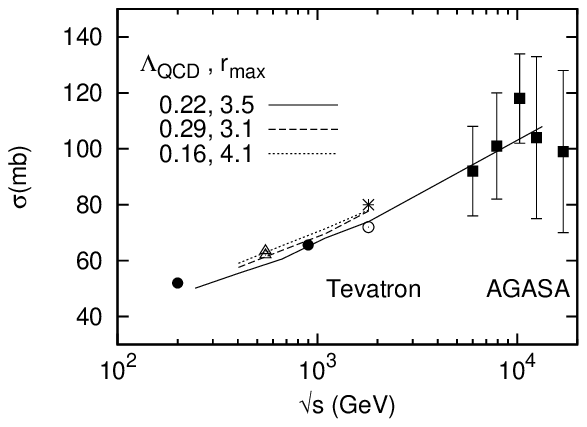,width=0.6\linewidth} 
  \caption{\label{fig:pptotrmaxlqcd} The total $pp$ cross section as a
    function of collision energy for various values of $r_{max}$ and
    $\Lambda_{QCD}$ in units of GeV$^{-1}$ and GeV respectively.  } }

The probability for a three-body system to contract to a single point
should be zero, and we have therefore also tested a wavefunction where 
small $r$-values are suppressed but where $\langle r^2\rangle$ has the same value. 
We then find essentially the same results for total cross sections,
but the reduced fluctuations imply that the cross sections for elastic
scattering and single diffractive excitation become larger.
This feature will be further discussed in sections \ref{sec:fluctwave}
and \ref{sec:res}.

\subsection{Dipole swing}
\label{sec:swing}

One problem with Mueller's model is the fact that saturation effects
are not included inside the individual dipole cascades. Thus only
pomeron loops which are cut in the particular Lorentz frame used are
taken into account, and the result is not frame independent. 
Non-linearities due to multiple interactions are included,
but the evolution of the dipole cascades obey the linear BFKL
equation.

We argued in ref.~\cite{Avsar:2006jy} that the missing saturation
effects can be taken into account by including the so called dipole
swing in the evolution. Although there is no analytical proof that
this would give a frame independent formalism, numerical simulations
in a MC implementation showed that the resulting evolution is
approximately frame independent.  We will show this in more detail in
subsection \ref{sec:frame}.

The swing is a process in which two dipoles $(\pmb{x}_i,\pmb{y}_i)$
and $(\pmb{x}_j,\pmb{y}_j)$ are replaced by two new dipoles
$(\pmb{x}_i,\pmb{y}_j)$ and $(\pmb{x}_j,\pmb{y}_i)$. The process can
be interpreted in two ways. The quark at $\pmb{x}_i$ and the antiquark
at $\pmb{y}_j$ form a colour singlet with probability $1/N_c^2$. In this
case the best approximation of the quadrupole field ought to be obtained
by the closest charge--anti-charge combinations.
Here the swing is therefore naturally suppressed by
$N_c^2$, and it should be more likely to replace two given dipoles
with two smaller ones. Secondly, we may see it as the result of a
gluon exchange between the dipoles, which results in a change in the
colour flow. In this case the swing would be proportional to
$\as^2$, which again is formally suppressed by $N_c^2$, compared
to the splitting process in \eqref{eq:dipkernel1}, which is
proportional to $\bar{\alpha}=N_c \as/\pi$.

In the MC implementation each dipole is randomly given one of $N_c^2$
possible colour indices. Only dipoles with the same colour index can swing, and
the weight for a swing
$(\pmb{x}_1,\pmb{y}_1),(\pmb{x}_2,\pmb{y}_2)\rightarrow
(\pmb{x}_1,\pmb{y}_2), (\pmb{x}_2,\pmb{y}_1)$ is determined by factor
proportional to
\begin{equation}
\frac{(\pmb{x}_1-\pmb{y}_1)^2 (\pmb{x}_2-\pmb{y}_2)^2}
{(\pmb{x}_1-\pmb{y}_2)^2 (\pmb{x}_2-\pmb{y}_1)^2}.
\end{equation}
This implies that the swing favours the formation of smaller dipoles.
The number of dipoles is not reduced by the swing, but the fact that
smaller dipoles have smaller cross sections gives the desired
suppression of the total cross section.

The swing is instantaneous in $Y$ for both the colour multipole and
gluon exchange interpretations. It is therefore not a vertex in the
sense of the dipole splitting whose probability is proportional to
$\Delta Y$.  In the MC implementation the swing is formulated as if it
was proportional to $\Delta Y$, but its strength is adjusted so that
it is effectively instantaneous. In ref.~\cite{Avsar:2007xh} \done{ref
  till Emil} it is shown that combining the dipole splitting and the
dipole swing, one can reproduce all colour correlations induced from
the multiple dipole interactions. 
In case each dipole is
restricted to single scattering, one can combine a splitting with one
swing at a time to reproduce all correlations, but without this
restriction, the maximum number of simultaneous swings needed in
combination with a splitting, for a system consisting of $N$ dipoles,
is $N-1$.

\subsection{Consistent treatment of confinement}
\label{sec:conf}

As the dipole model is formulated within perturbative QCD, confinement
effects are naturally not included. Obviously, one cannot let the
dipoles in the cascade become too large, and it is then natural to
introduce a scale, such as our $r_{max}$ parameter, so that large
dipoles are suppressed.

Similarly, the scattering of dipoles is calculated perturbatively, and
the interaction range is therefore longer than what we would expect
from confinement.  The formula for $f_{ij}$ in \eqref{eq:dipamp} is
just the two dimensional Coulomb potential which for large distances
behaves as
\begin{eqnarray}
f_{ij} \sim \frac{\as^2}{8} \frac{(\pmb{x}_i-\pmb{y}_i)^2
(\pmb{x}_j-\pmb{y}_j)^2}{\pmb{b}^4},
\end{eqnarray}
where $\pmb{b}=\frac{1}{2}((\pmb{x}_i+\pmb{y}_i)-(\pmb{x}_j+\pmb{y}_j))$ 
is the impact parameter of the dipole--dipole collision. 
Thus the scattering probability falls off only as a power of $\pmb{b}$, 
and not as an exponential as one would expect from a confining field. 
The expression for $f_{ij}$ can be written as 
\begin{eqnarray}
  f(\pmb{x}_i,\pmb{y}_i|\pmb{x}_j,\pmb{y}_j) =
  \frac{g^4}{8} ( \Delta(\pmb{x}_i - \pmb{x}_j)
  - \Delta(\pmb{x}_i - \pmb{y}_j) - \Delta(\pmb{y}_i - \pmb{x}_j) +
  \Delta(\pmb{y}_i - \pmb{y}_j) )^2\label{eq:ddgreen}
\end{eqnarray}
where $\Delta(\pmb{r})$ is the Green's function given by 
\begin{eqnarray}
  \Delta(\pmb{r}) = \int \frac{d^2\pmb{k}}{(2\pi)^2}
  \frac{e^{i\pmb{k} \cdot\pmb{r}}}{\pmb{k}^2}.
\label{eq:coulombprop}
\end{eqnarray}
 
To take confinement into account we can replace the infinite range
Coulomb potential with a screened Yukawa potential. This implies that
the Coulomb propagator $1/\pmb{k}^2$ in \eqref{eq:coulombprop} is
replaced by $1/(\pmb{k}^2+M^2)$, where $M=1/r_{max}$ is the
confinement scale.  In this case the integral in
\eqref{eq:coulombprop} is replaced by
\begin{eqnarray}
\int \frac{d^2\pmb{k}}{(2\pi)^2} \frac{e^{i\pmb{k} \cdot\pmb{r}}}{\pmb{k}^2 + M^2} 
= \frac{1}{2\pi}K_0(rM)
\end{eqnarray} 
where $K_0$ is a modified Bessel function. The expression in
\eqref{eq:dipamp} is then replaced by
\begin{eqnarray}
f_{ij} \to \frac{\alpha_s^2}{2} \biggl ( K_0(|\pmb{x}_i-\pmb{x}_j|/r_{max}) - 
K_0(|\pmb{x}_i-\pmb{y}_j|/r_{max}) - 
\nonumber \\- K_0(|\pmb{y}_i-\pmb{x}_j|/r_{max}) + 
K_0(|\pmb{y}_i-\pmb{y}_j|/r_{max}) \biggr )^2.
\label{eq:conffij}
\end{eqnarray} 
For small separations where $r << r_{max}$, the function
$K_0(r/r_{max})$ behaves like ln$(r_{max}/r)$ and we then immediately
get the result in \eqref{eq:dipamp}.  For large separations where $r
>> r_{max}$, $K_0(r/r_{max})$ falls off exponentially $\sim
\sqrt{\frac{\pi r_{max}}{r}}e^{-r/r_{max}}$, as expected from
confinement.

To be consistent we should then also modify the dipole splitting
kernel accordingly. The dipole splitting probability in
\eqref{eq:dipkernel1} can be written in the form
\begin{eqnarray}
\frac{d\mathcal{P}}{dY}=\frac{\bar{\alpha}}{2\pi}d^2\pmb{z}
\frac{(\pmb{x}-\pmb{y})^2}{(\pmb{x}-\pmb{z})^2 (\pmb{z}-\pmb{y})^2} = 
\frac{\bar{\alpha}}{2\pi}d^2\pmb{z}\biggl ( \frac{\pmb{x}-\pmb{z}}
{(\pmb{x}-\pmb{z})^2} - \frac{\pmb{y}-\pmb{z}}{(\pmb{y}-\pmb{z})^2}\biggr )^2.
\label{eq:dipkernel}
\end{eqnarray}
The two terms in this expression are each obtained from the
integration
\begin{eqnarray}
\int \frac{d^2\pmb{k}}{(2\pi)^2i} \frac{\pmb{k}e^{i\pmb{k} \cdot\pmb{r}}}
{\pmb{k}^2} = -\pmb{\nabla}\int \frac{d^2\pmb{k}}{(2\pi)^2} 
\frac{e^{i\pmb{k} \cdot\pmb{r}}}{\pmb{k}^2}.
\end{eqnarray}
Once again making the change $1/\pmb{k}^2\to 1/(\pmb{k}^2+M^2)$, and
noting that $\pmb{\nabla}K_0(r/r_{max}) = - \frac{\pmb{r}}{r\cdot
  r_{max}}K_1(r/r_{max})$, we may replace \eqref{eq:dipkernel} by
\begin{eqnarray}
\frac{d\mathcal{P}}{dY} \to \frac{\bar{\alpha}}{2\pi}d^2\pmb{z} \biggl (
\frac{1}{r_{max}}\frac{\pmb{x}-\pmb{z}}{|\pmb{x}-\pmb{z}|}K_1(|\pmb{x}-\pmb{z}|/r_{max}) - 
\frac{1}{r_{max}}\frac{\pmb{y}-\pmb{z}}{|\pmb{y}-\pmb{z}|}K_1(|\pmb{y}-\pmb{z}|/r_{max})
\biggr )^2. \nonumber \\
\label{eq:moddipkernel}
\end{eqnarray}
For small arguments $K_1(r/r_{max}) \approx \frac{r_{max}}{r}$, from
which we get back the result in \eqref{eq:dipkernel}, while for large
arguments $K_1(r/r_{max}) \sim \sqrt{\frac{\pi
r_{max}}{r}}e^{-r/r_{max}}$, and once again we obtain an
exponentially decaying field.

\subsection{Frame independence}
\label{sec:frame}

We will in this subsection demonstrate the frame independence by
showing some explicit results obtained using the MC implementation.

The cross section obtained when the right-moving (left-moving) cascade
is evolved a rapidity distance $Y_0$ ($Y-Y_0$) is denoted
$\sigma(Y_0,Y)$, and in figure \ref{fig:ppframeindp} we show the
relative difference $\Delta \sigma/\sigma = (\sigma(Y_0,Y) -
\sigma(Y/2,Y))/\sigma(Y/2,Y)$ plotted \vs\ $\delta =Y_0/Y$.  The
figure shows results both including the dipole swing and without the
swing. Without the swing the cross section is too large when $\delta
\to 0$ or $\delta \to 1$. As expected, the degree of frame dependence
is increasing for larger $\sqrt{s}$, when the saturation effects
within cascades become more important.  When we include the swing, we
see that the cross section is (within errors) independent of the
Lorentz frame used.

\FIGURE[t]{
  \epsfig{file=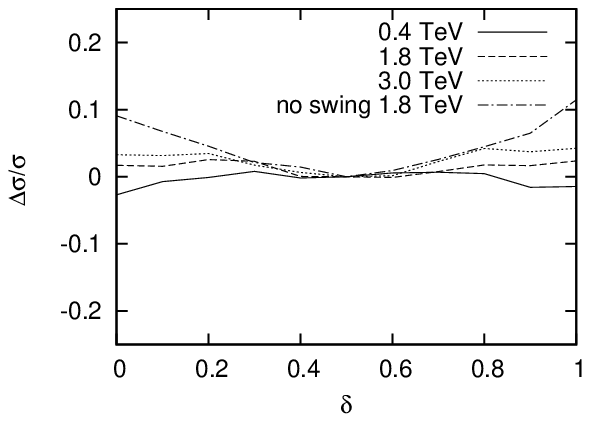,width=0.6\linewidth} 
  \caption{\label{fig:ppframeindp} The quantity $\Delta \sigma/\sigma
    = (\sigma(Y_0, Y) - \sigma(Y/2, Y))/\sigma(Y/2, Y)$ plotted as a
    function of $\delta=Y_0/Y$ for different collision energies: full
    line $\sqrt{s}=0.4$~TeV, dashed line $\sqrt{s}=1.8$~TeV and dotted
    line $\sqrt{s}=3.0$~TeV. Also shown is the result excluding the
    dipole swing for $\sqrt{s}=1.8$~TeV (dash-dotted line).  } }

\FIGURE{
  \epsfig{file=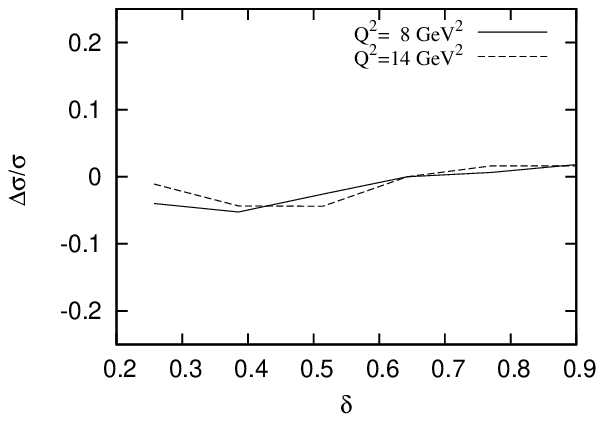,width=0.6\linewidth} 
  \caption{\label{fig:gpframeindp} The quantity $\Delta \sigma/\sigma
    = (\sigma(Y_0,Y)) - \sigma(0.64Y,Y)/\sigma(0.64Y,Y)$, plotted as
    a function of $\delta=Y_0/Y$ in DIS for $W=220$~GeV and
    $Q^2=8$~GeV$^2$ (solid line) and $Q^2=14$~GeV$^2$ (dashed line).}}

\FIGURE[t]{
  \epsfig{file=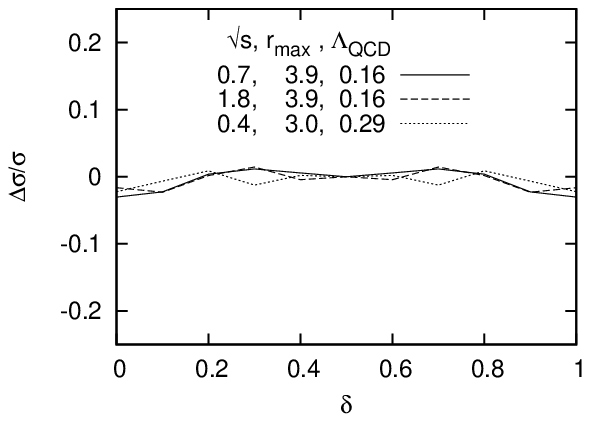,width=0.6\linewidth} 
  \caption{\label{fig:ppframeindp2} The quantity $\Delta
    \sigma/\sigma$ (as defined in figure \ref{fig:ppframeindp}) as a
    function of $\delta$ for different values of collision energy,
    $r_{max}$ and $\Lambda_{QCD}$ (in units of TeV, GeV$^{-1}$ and GeV
    respectively) in $pp$ scattering.  } 
}

\FIGURE[]{
  \epsfig{file=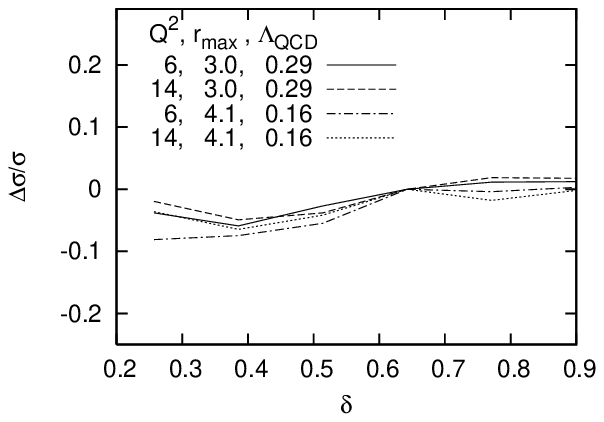,width=0.6\linewidth} 
  \caption{\label{fig:gpframeindp2} The quantity $\Delta
    \sigma/\sigma$ (as defined in figure \ref{fig:gpframeindp}) as a
    function of $\delta$ in DIS for $W=220$~GeV for different values of
    $Q^2$, $r_{max}$ and $\Lambda_{QCD}$ (in units of GeV$^2$,
    GeV$^{-1}$ and GeV respectively).  } }

Figure \ref{fig:gpframeindp} shows similar results for DIS.  Here the
cross section is not exactly frame independent, with a tendency of
getting larger as we give more of the total rapidity interval to the
evolution of the photon. (This may not be very clear from the figures 
shown but it can be seen more clearly for dipole-proton scattering 
where we do not have ambiguities in choosing the $Y$ interval as in DIS.)
There seems to be two causes for this
behaviour, and neither of them are related to saturation or the dipole
swing. The first cause is the running of the coupling, especially when
the dipole--dipole scattering amplitudes are calculated (see next
subsection). The second cause is the treatment of energy conservation
in the dipole--dipole scattering. In case we use a fixed coupling when
calculating the scattering amplitude (but still using a running
coupling in the evolution), the result appears to be more frame
independent.  These issues need to be handled more carefully, and we
intend to look at them in forthcoming publications.  Note however that
the difference is less than 10 percent for both $Q^2$ values, and it
is also not dependent on the energy.  As our model is not expected to
have a better accuracy than this, we can consider the model to be
essentially frame independent also for DIS.

In figures \ref{fig:ppframeindp2} and \ref{fig:gpframeindp2} we also 
show the same diagrams for different values of the parameters 
$r_{max}$ and $\Lambda_{QCD}$, and we can see a similar behaviour 
as in the previous figures. The frame independence of the model 
is therefore not dependent on the precise values of these parameters. 

Naturally also the elastic cross section must be frame independent.
Diffraction and elastic scattering will be studied in sections
\ref{sec:diff} and \ref{sec:res}, and the numerical result for
$\sigma_{el}$ is indeed independent of the Lorentz frame used.


\subsection{Running Coupling}
\label{sec:running}

In our simulations we use a running \as\ both in the dipole
splitting and in the dipole--dipole scattering probability. Recent NLO
studies of the dipole evolution
\cite{Balitsky:2006wa,Kovchegov:2006wf} have revealed a fairly
complicated structure for the running of \as. However, in \cite{Balitsky:2007}
this was shown to simplify in the strongly ordered limits, implying that the
relevant scale in the dipole splitting is determined by $\min(r, r_1,
r_2)$, where $r$ is the mother dipole which splits into $r_1$ and
$r_2$. This is also the scale we have been using in our simulations.

For the dipole--dipole scattering, the situation is more complicated.
We have two powers of \as\ and there are six different scales 
involved. With the two colliding dipoles $(\pmb{x}_i,\pmb{y}_i)$ and
$(\pmb{x}_j,\pmb{y}_j)$ we have besides, their sizes, also the
distances $|\pmb{x}_i-\pmb{x}_j|$, $|\pmb{x}_i-\pmb{y}_j|$,
$|\pmb{y}_i-\pmb{x}_j|$ and $|\pmb{y}_i-\pmb{y}_j|$ (cf.\
\eqref{eq:ddgreen}). Here we have tried two alternatives. In the 
first case the scale of \as\ for both powers is determined by 
$\min(|\pmb{x}_i - \pmb{y}_i|,|\pmb{x}_j-\pmb{y}_j|, 
|\pmb{x}_i-\pmb{y}_j|, |\pmb{y}_i-\pmb{x}_j|)$. In the second 
case, we 
associate one \as\ to each of the colliding dipoles and choose the
scales $\min(|\pmb{x}_i - \pmb{y}_i|,|\pmb{x}_i-\pmb{x}_j|,
|\pmb{x}_i-\pmb{y}_j|, |\pmb{y}_i-\pmb{x}_j|, |\pmb{y}_i-\pmb{y}_j|)$
and $\min(|\pmb{x}_j - \pmb{y}_j|,|\pmb{x}_i-\pmb{x}_j|,
|\pmb{x}_i-\pmb{y}_j|, |\pmb{y}_i-\pmb{x}_j|, |\pmb{y}_i-\pmb{y}_j|)$
respectively. To avoid divergencies, \as\ is frozen below
the scale $p_\perp=2/r_{\max}$. 

It turns out that the degree of frame dependence in $\gamma^*p$ is
similar in both cases and the results presented in this paper have all
been calculated using the first alternative.

\section{Diffractive and elastic scattering in the dipole model}
\label{sec:diff}

\subsection{Formalism}
\label{sec:formalism}

In this section we will describe the formulas we are going to use in
calculating the various diffractive and elastic cross sections. We
shall rely on the dipole version of the Good and Walker picture of
diffraction \cite{Good:1960ba} where the scattering eigenstates are
given by the dipole states. The identification of the QCD parton
states as the eigenstates of diffraction is due to the work of
Miettinen and Pumplin \cite{Miettinen:1978jb}. The situation is
complicated by the fact that the states of the proton or the virtual
photon depend on the Lorentz frame used, and we will here quite
closely follow the formalism presented in \cite{Hatta:2006hs}.
 
In the Good and Walker picture of diffraction there is a normalized
and complete set of real particle states $\{|N\rangle\}$ with fixed
quantum numbers. In addition we have eigenstates of the scattering,
$\{|n\rangle \}$, which also form a complete set of normalized states.
Assume that we have two incoming particles, one right-moving particle
$|R\rangle$ and one left-moving particle $|L\rangle$. These particles
can then be diffracted onto the various particle states
$\{|N\rangle\}$ and $\{|M\rangle\}$,
which 
carry the quantum numbers of $|R\rangle$ and $|L\rangle$ respectively. The 
incoming wave is given by 
\begin{eqnarray}
|\psi_I\rangle = |R,L\rangle = \sum_{n,m} c_n^Rc_m^L |n,m\rangle. 
\end{eqnarray}
The scattered wave is obtained by operating with Im$\pmb{T}$ on
$|\psi_I\rangle$, where $\pmb{T}$ is the scattering operator. It reads
\begin{eqnarray}
|\psi_S\rangle = \mathrm{Im}\pmb{T}|\psi_I\rangle = \sum_{n,m}c_n^Rc_m^L
t(n,m) |n,m\rangle.
\end{eqnarray}
The probability for diffractive scattering is given by 
\begin{equation}
\langle \psi_S|\psi_S\rangle = \sum_{n,m}  P_n^R P_m^L [t(n,m)]^2 = 
\langle t^2 \rangle_{R,L}.
\label{eq:sigmadiff}
\end{equation}
We have here identified $|c_n|^2 \equiv P_n$ for both $R$ and $L$ with
the probability distribution for the dipole configurations inside the
particles. Note that the sum $\sum_n$ actually involves a sum over the
dipole occupation number $n$ as well as integrations over the
transverse coordinates of each dipole, together with sums over their
colour and spin configurations.
 
Using the completeness of the states $\{|N,M\rangle\}$, the expression
in \eqref{eq:sigmadiff} can be written in the following form:
\begin{eqnarray}
\langle \psi_S|\psi_S\rangle = |\langle R, L|\psi_S\rangle|^2 + 
\sum_{N\neq R} |\langle N, L|\psi_S\rangle|^2 + \sum_{M\neq L} |\langle R, M|\psi_S\rangle|^2 +
\nonumber \\
+ \sum_{N\neq R}\sum_{M\neq L} |\langle N, M|\psi_S\rangle|^2.
\label{eq:GWdiff}
\end{eqnarray}
Here the first term on the RHS corresponds to elastic scattering,
where both $R$ and $L$ emerge intact from the collision. The second
(third) piece gives the probability of the excitation of $R$ ($L$)
into one of the states $N$ ($M$) with $L$ ($R$) remaining intact. This
corresponds to single diffractive excitation. Finally, the last term
takes into account the fact that \emph{both} $R$ and $L$ may transform
into excited states $N$ and $M$, which thus corresponds to double
diffractive excitation.

We note that the different terms in \eqref{eq:GWdiff} correspond to
different averages of $t(n,m)^2$. The sum of the single diffractive
excitation and the elastic cross section can be calculated as follows
\begin{eqnarray}
\sum_{N} |\langle N, L|\psi_S\rangle|^2 &=& \sum_{N} \biggl | \sum_{n,m} 
c_n^{N*}\, c_n^R\, P_m^L\, t(n,m) \biggr |^2 = \sum_{N} \sum_{n,n'} c_n^{N*} \, 
c_n^R \, c_ {n'}^N\, c_{n'}^{R*} \langle t(n) \rangle_L^2 \nonumber \\
&=& \sum_n P_n^R  \langle t(n) \rangle_L^2 = \langle \langle t \rangle_L^2 
\rangle_R
\label{eq:GWsd}
\end{eqnarray}
where we used the completeness of the states $\{N\}$, 
\begin{eqnarray}
\sum_N c_n^{N*} \, c_{n'}^N  = \delta_{nn'}.
\end{eqnarray}

Each of the coefficients $c_n$ above is to be evaluated at a certain
rapidity $Y_0$.  The total rapidity interval between $R$ and $L$ is
determined by the total cms energy $\sqrt{s}$ of the process. For $pp$
scattering $Y$ is simply given by ln$(s/M_p^2)$, while the situation
is a bit more subtle in DIS. How we determine $Y$ in DIS was discussed
above in section \ref{sec:photon}.

In Mueller's dipole model the scattering amplitude $t(n,m)$ is given
by the eikonal form $1 - e^{-F}$, where $F=\sum_{ij} f_{ij}$ is
defined in \eqsref{eq:dipamp}, (\ref{eq:T}).  The different
contributions to the diffractive cross section in \eqref{eq:GWdiff}
are then given by
\begin{eqnarray}
  \frac{d\sigma_{el}}{d^2\pmb{b}} &=&
  \biggl \langle  1 - e^{-F} \biggr \rangle_{R,L}^2 \label{eq:sigmael} \\
  \frac{d\sigma_{SD}^R}{d^2\pmb{b}} &=&
  \biggl \langle \biggl \langle 1 - e^{-F} 
  \biggr \rangle_L^2 \biggr \rangle_R -
  \biggl \langle  1 - e^{-F} \biggr \rangle_{R,L}^2 \label{eq:sigmasdr} \\
  \frac{d\sigma_{SD}^L}{d^2\pmb{b}} &=& \biggl \langle \biggl \langle 1 - e^{-F} 
  \biggr \rangle_R^2 \biggr \rangle_L - \biggl \langle
  1 - e^{-F} \biggr \rangle_{R,L}^2 \label{eq:sigmasdl} \\
  \frac{d\sigma_{DD}}{d^2\pmb{b}} &=& \biggl \langle \biggl (1 - e^{-F} \biggr )^2 \biggr 
  \rangle_{R,L} - \biggl \langle \biggl \langle 1 - e^{-F} 
  \biggr \rangle_L^2 \biggr \rangle_R
  - \biggl \langle \biggl \langle 1 - e^{-F} \biggr \rangle_R^2 \biggr \rangle_L
  \nonumber \\ &+& 
  \biggl \langle  1 - e^{-F} \biggr \rangle_{R,L}^2. 
  \label{eq:sigmadd}
\end{eqnarray}
Here $\sigma_{SD}^R$ ($\sigma_{SD}^L$) is the cross section for the
diffractive excitation of $R$ ($L$). Similarly $\sigma_{el}$ and
$\sigma_{DD}$ stand for the elastic and double diffractive cross
sections respectively.  Summing these four contributions we get the
total diffractive cross section
\begin{eqnarray}
  \frac{d\sigma_{diff}}{d^2\pmb{b}} =
  \langle \psi_S|\psi_S\rangle = \sum_{n,m} P_n^R \,P_m^L \,[t(n,m)]^2 = 
  \biggl \langle \biggl (1 - e^{-F} \biggr )^2 \biggr \rangle_{R,L}.
\label{eq:sigmadiffb} 
\end{eqnarray}

Assume now that the state $R$ is evolved up to $Y_0$ while $L$ is
evolved up to $Y-Y_0$, with $Y$ the total rapidity interval.  The
total and elastic cross sections given by
\begin{eqnarray}
\sigma_{tot}(Y) = 2 \int d^2\pmb{b} \,\biggl \langle 1 - e^{-F(\pmb{b})} 
\biggr \rangle_{R,L}\,\,\, \mathrm{and}\,\,\, 
\sigma_{el}(Y) = \int d^2\pmb{b} \,\biggl \langle 1 - e^{-F(\pmb{b})}
\biggr \rangle_{R,L}^2
\label{eq:sigmatot}
\end{eqnarray}
are necessarily independent of $Y_0$ due to the requirement of frame
independence.

The diffractive cross section in \eqref{eq:sigmadiffb} is, however,
\emph{not} independent of $Y_0$.  This expression gives the
probability for diffraction at a particular value of $Y_0$,
\ie\ the chance that we find a rapidity gap around that
particular $Y_0$. If we calculate \eg\ $\sigma_{SD}^R$ at a
specific $Y_0$, we obtain the cross section where the diffractively
excited right-moving particle is confined within the rapidity range
$(0, Y_0)$.  This is approximately equivalent to a maximal diffractive
mass given by $M_{X,max}^2 \approx e^{Y_0}$.  Taking the derivative
with respect to $Y_0$ therefore gives the mass distribution
$d\sigma_{SD}^R/d Y_0 = d\sigma_{SD}^R/d \ln(M_X^2)$.

For the total diffractive cross section in \eqref{eq:sigmadiffb}
we thus expect, in the case of the symmetric $pp$ collision, a maximum
when $Y_0=Y/2$, and a decrease when either $Y_0 \to 0$ or $Y_0 \to Y$.
For asymmetric scattering as in DIS, we expect that it is easier to
excite the photon.  For a right-moving photon the diffractive cross
section should therefore be smallest when $Y_0 \to 0$. This discussion
is in accordance with the analysis by Hatta et al.
\cite{Hatta:2006hs}. In this reference the diffractive excitation of
the proton is neglected, while in our formalism we can also take
the proton excitation into account.

The results obtained for pp collisions and DIS are presented in
section \ref{sec:res}.








\subsection{Importance of fluctuations}
\label{sec:fluctuations}

The impact of fluctuations upon the small-$x$ evolution has gathered
considerable interest lately.
As mentioned above, the various expressions for the cross sections in
formulas \eqsref{eq:sigmael}-(\ref{eq:sigmadd}) are all obtained by
taking different averages of the quantity $(1-e^{-F})^2$.  The
diffractive excitation is therefore completely determined by the
fluctuations in the colliding systems and the interaction
probabilities.

\subsubsection{Different sources}
\label{sec:sources}

There are several sources of fluctuations in the various expressions
in \eqsref{eq:sigmael}-(\ref{eq:sigmadd}), related to variations in
impact parameter, in the dipole cascades, and in initial the wave functions for 
the photons and protons.  Many analyses include part of the
fluctuations, assuming this to give the dominant contribution.  Thus
the dipole-saturation model by Golec-Biernat and W\"usthoff
\cite{Golec-Biernat:1998js,Golec-Biernat:1999qd} takes into account
fluctuations in the photon wave function and the emission of the first
gluon in the photon cascade, while the model of Kowalski and Teaney
(KT) \cite{Kowalski:2003hm} emphasizes the fluctuations in the impact
parameter. Hatta et al.\ \cite{Hatta:2006hs} includes the fluctuations
in the photon cascade, but assumes that the fluctuations in the proton
cascade can be neglected. As a result fits to data can give different
results for the impact parameter profile, and different approaches can
give different ratios for the elastic cross section and diffractive
excitation. As an illustration we will here compare the fluctuations
in our model with those in the Kowalski--Teaney model.

In the KT model the differential dipole--proton cross section in impact
parameter space is given by the eikonal
\begin{eqnarray}
\frac{d\sigma_{dp}}{d^2\pmb{b}} = 2(1 - e^{-\Omega(r, b)/2}),
\end{eqnarray}
where $r$ is the dipole size. The opacity $\Omega$ is modeled by a factorized form
\begin{eqnarray}
\Omega(r, b) = \frac{\pi^2}{N_c}r^2 \as(\mu^2)xg(x,\mu^2)T(b),
\label{eq:kowomega}
\end{eqnarray}
where $T(b)$ is the transverse profile function of the proton. In
$\Omega(r, b)$ there is also a dependence on $W^2 \propto 1/x$ which
is omitted here. To determine the impact parameter profile it is assumed
that the $t$ dependence of the diffractive vector meson production
cross section is given by an exponential, which in turn implies a
Gaussian profile for $T(b)$. The two unknown parameters of this
Gaussian are then determined by a fit to diffractive $J/\psi$
production data.

For a virtual photon the only fluctuations are those in the dipole
size, $r$, and impact parameter, $b$.  The diffractive cross section
is calculated as
\begin{eqnarray}
\sigma_{diff}^{KT} = \int d^2\pmb{b} \int d^2\pmb{r} \, dz\, \psi_\gamma(r,z,Q^2) \,\biggl 
( 1 - e^{-\Omega (r, b)/2} \biggr )^2,
\end{eqnarray}
where $\psi_\gamma(r,z,Q^2) = |\psi_T(r,z,Q^2)|^2+|\psi_L(r,z,Q^2)|^2$.  

When comparing the two models it may seem natural to compare
$\Omega(r, b)/2$ with the average $\langle F\rangle_{d,p}$, where
$\langle \rangle_{d,p}$ denotes the averaging over the dipole and
proton cascades and the initial proton wavefunction. The corresponding total
and diffractive cross sections would then read
\begin{eqnarray}
\sigma_{tot} &=& 2 \int d^2\pmb{b} \int d^2\pmb{r} \, dz \, \psi_\gamma(r,z,Q^2) 
\, \biggl (1 - e^{-\langle F(\pmb{r}, \pmb{b})\rangle_{d,p}}\biggr ) , \label{eq:kowtot1}\\
\sigma_{diff}^{(1)} &=& \int d^2\pmb{b} \int d^2\pmb{r} \, dz \, \psi_\gamma(r,z,Q^2) \,
\biggl ( 1 - e^{-\langle F(\pmb{r}, \pmb{b})\rangle_{d,p}} \biggr )^2.
\label{eq:kowdiff1}
\end{eqnarray} 

This is, however, not necessarily correct. The opacity $\Omega$ is in
the KT model determined by a fit to data for the total cross section.
What is directly determined is therefore $1 - e^{-\Omega/2}$,
rather than $\Omega$ itself. Thus a more direct analogy to our
model would be the quantity $\langle 1 - e^{-F} \rangle$, which would
give the same total cross section as \eqref{eq:sigmatot} and imply the
following form for the diffractive cross section:
\begin{eqnarray}
\sigma_{diff}^{(2)} = \int d^2\pmb{b} \int d^2\pmb{r} \, dz\, \psi_\gamma(r,z,Q^2) 
\,\biggl \langle 1 - e^{-F(\pmb{r}, \pmb{b})}
 \biggr \rangle_{d,p}^2.
\label{eq:kowdiff2}
\end{eqnarray}
Note that this is not the same as the elastic contribution in
\eqref{eq:sigmael} since in that case also the photon wave function is
included in the squared average (\eqref{eq:sigmael} is meaningless in
DIS since the virtual photon cannot scatter elastically).

The expressions in \eqref{eq:kowdiff1} and \eqref{eq:kowdiff2} should
be compared with the results in our model, obtained by integrating
\eqref{eq:sigmadiffb} over $\pmb{b}$:
\begin{eqnarray}
\sigma_{diff} = \int d^2\pmb{b} \int d^2\pmb{r} \, dz\, \psi_\gamma(r,z,Q^2) 
\,\biggl \langle \biggl ( 
1 - e^{-F(\pmb{r}, \pmb{b})} \biggr )^2 \biggr \rangle_{d,p}.
\label{eq:sigmadiff2}
\end{eqnarray}
Such a comparison is interesting as a way to gauge the role played by
the fluctuations.  \FIGURE[t]{
  \epsfig{file=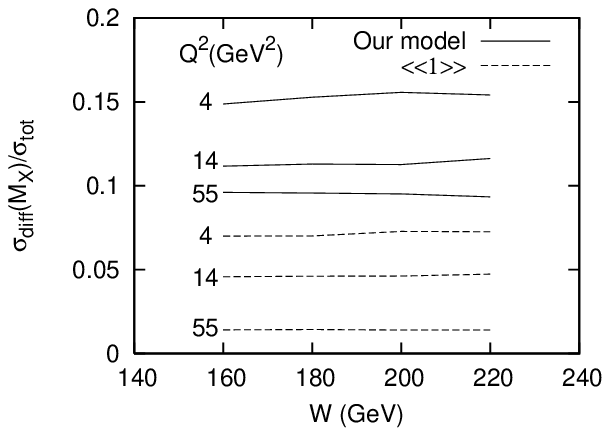,width=0.5\linewidth}%
  \epsfig{file=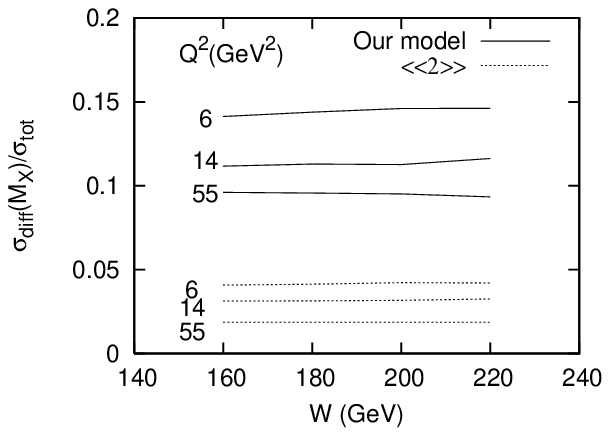,width=0.5\linewidth}
  \caption{\label{fig:gpfluct} The ratio of the diffractive cross
    section to the total cross section for $M_X< 32$~GeV$^2$. Our
    results obtained from \eqref{eq:sigmadiff2} (full lines) are
    compared to results obtained from both \eqref{eq:kowdiff1}, marked
    $\langle\langle1\rangle\rangle$ in the left figure (dashed lines),
    and \eqref{eq:kowdiff2}, marked $\langle\langle2\rangle\rangle$ in
    the right figure (dotted lines).  The total cross section is
    calculated according to \eqref{eq:sigmatot}.  } }

The results obtained from \eqref{eq:kowdiff1} and \eqref{eq:kowdiff2}
are shown in figure \ref{fig:gpfluct}, together with the results
obtained from \eqref{eq:sigmadiff2} which includes all fluctuations.
We immediately notice the very large effects in our model from the
fluctuations in the cascades and the proton wave function. The
diffractive cross section calculated from \eqref{eq:kowdiff1} is seen
to be a factor 2-3 below the result obtained from
\eqref{eq:sigmadiff2}, while the result from \eqref{eq:kowdiff2} is
around a factor 4 lower than \eqref{eq:sigmadiff2}.

We conclude that in our model a large fraction of the fluctuations
determining the diffractive cross sections is caused by the dipole
cascade evolutions. In order to obtain a similar result in the KT
model it is therefore necessary to have larger fluctuations due to the
impact parameter dependence, which means an impact parameter profile
which is more narrow, \ie\ more black and white compared to Mueller's
dipole cascade model, where the average scattering can be ``grey''
overall, since the fluctuations in the cascades means that some events
are almost black while other are almost white. This can clearly be seen
in figure \ref{fig:bprofile} where we compare the impact parameter
profile from the KT model to that obtained from our model. (A similar 
effect, although less pronounced, is observed in the profile
for $pp$ scattering presented in ref.~\cite{Avsar:2006jy}. Also here
the profile obtained in our model has a somewhat higher tail for
large impact parameters than the Gaussian fit to Tevatron data by 
Sapeta and Golec-Biernat \cite{Sapeta:2005ba}.)
 
\FIGURE[t]{
  \epsfig{file=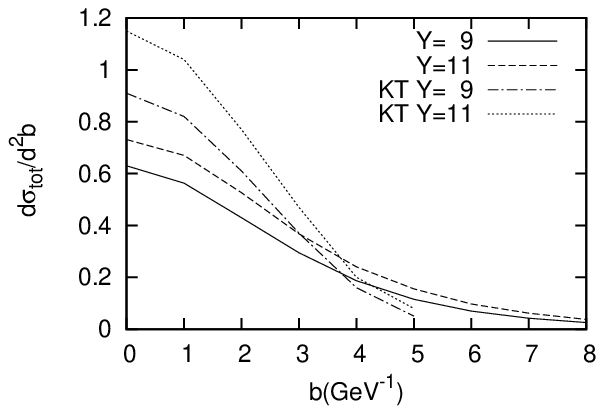,width=0.6\linewidth} 
  \caption{\label{fig:bprofile} The impact parameter profile of
    dipole--proton collisions for an initial dipole size $r=2
    \mathrm{GeV}^{-1} \approx 0.4 \mathrm{fm}$ at two different
    energies, $Y=\ln{s}=9$ and 11. Our results (solid and dashed
    lines) are compared to those from the Kowalski--Teaney (KT)
    \cite{Kowalski:2003hm} model
    (dot-dashed and dotted lines).  } }

\subsubsection{Wave functions}
\label{sec:fluctwave}

\emph{Photon}

The photon wave functions in \eqref{eq:psigamma} for longitudinal and
transverse photons are fully determined by perturbation theory.
We note that these wave
functions are not normalized, even for real photons. This is, however,
not in contradiction with the assumption that the states
$\{|N\rangle\}$ are normalized. The generic photon state can be
written
\begin{eqnarray}
  |\gamma (Y_0)\rangle = c_0^\gamma(Y_0)\,|\gamma_d\rangle + 
  \sum_{n= 1}^\infty c_n^\gamma(Y_0) \, |n\rangle,
\end{eqnarray}  
where $|\gamma_d\rangle$ is the component of the photon coupling
directly to the quarks. While the state $|\gamma\rangle$ is
normalized, the two separate components above are not. Rescattering of
the component $|\gamma_d\rangle$ can be neglected, as it is
proportional to $\alpha_{\mathrm{em}}$. It is therefore sufficient to only keep
the contribution from $|\gamma_h \rangle = \sum_n c_n |n\rangle$,
which is not a normalized state.

In DIS it is not meaningful to consider the elastic
$\gamma^* p$ scattering, as the virtual photon can never be
detected as a real particle. The closest analogies to elastic
scattering are given by Deeply Virtual Compton Scattering (DVCS) and
the exclusive reactions $\gamma^* p \rightarrow V p$, with $V$ a
vector meson. We will return to these processes in a future
publication.

\emph{Proton}

The wave function for the proton is much less well defined. The expressions
in \eqsref{eq:pwave1} or (\ref{eq:pwave2}) ought to be interpreted as
probability distributions rather than quantum mechanical wave functions,
which can be used to determine the interference effects present in 
\eqsref{eq:sigmael}-(\ref{eq:sigmadd}). The fluctuations in the wave
functions influence the terms in \eqsref{eq:sigmael}-(\ref{eq:sigmadd})
in which the average of $1-e^{-F}$ is taken before the square (\ie\ 
events containing an elastic proton).

As discussed in section \ref{sec:proton} a wave function where three
particles can simultaneously be in a single point is not realistic.
A Gaussian distribution in impact parameter is usually motivated
by the exponential dependence on $t$ for the elastic cross section.
However, the constant $t$-slope is (except possibly for the 
highest Tevatron energy) only valid for $|t|<0.15\, \mathrm{GeV}^2$,
corresponding to $b>2/\sqrt{0.15}\, \mathrm{GeV}^{-1} \approx 1\, \mathrm{fm}$,
and therefore a suppression for small $r$ is still compatible with this
constraint. Simulations with such a wave function reduces the fluctuations
and increases
the cross sections for elastic scattering and single diffractive 
excitation, while leaving the total and the total diffractive cross
sections unchanged, provided the average $\langle r^2 \rangle$ is kept the same.

There is, however, also a more fundamental problem with the proton wave function.
In the Good and Walker formalism the hadronic states $\{|N\rangle\}$ form a complete
set. This implies that before the cascade has started, there 
must be other hadron states with wavefunctions orthogonal to the proton 
wavefunction. This calls for a detailed dynamical scheme describing
the relevant degrees of freedom for the hadronic
states. With the approximation $M_X^2 \approx \exp{Y_0}$ these orthogonal states
also have the same mass as the proton, which increases the problem further.

Lacking a real quantum-mechanical description of the proton wave function,
we can still get an \emph{upper limit} for elastic scattering and single diffraction
by removing the contribution from the initial wave function fluctuations. This is
obtained if we integrate over the initial wave functions in 
\eqsref{eq:sigmael}-(\ref{eq:sigmadd}) after taking the squares.
Note that the average over different evolutions is still taken before the
square, and therefore the fluctuations in the cascade evolution and the impact
parameter dependence are still included. Note also that this does not affect
the result for the total cross section in \eqref{eq:sigmatot} or the total
diffractive cross section in \eqref{eq:sigmadiffb} (which also includes
the elastic cross section).

\subsubsection{Non-leading effects}
\label{sec:nonleading}

It was early pointed out by Mueller
and Salam \cite{Mueller:1996te} that there are extremely large fluctuations 
in the leading order cascade evolution. Expanding the exponential in $\langle 1 -
e^{-F} \rangle$ we have
\begin{eqnarray}
\biggl \langle 1 - e^{-F} \biggr \rangle = \sum_{k=1}^\infty \frac{(-1)^{k-1}}{k!}
\langle F^k \rangle.
\end{eqnarray} 
Here $\langle F^k \rangle$ could be interpreted as a contribution from
the exchange of $k$ pomerons, but such an interpretation may be difficult
 as it was numerically 
demonstrated\footnote{The numerical result was
  anticipated by Mueller \cite{Mueller:1994gb} who performed
  analytical calculations on a simple toy model in which transverse
  coordinates are neglected.} by Salam \cite{Salam:1995uy} that
$\langle F^k \rangle \sim (k!)^2$, and therefore this series is strongly
divergent. The reason for this is the existence of rare events with a
large number of dipoles and large values of $F$, which make $\langle
F^k \rangle$ blow up for large $k$. On the other hand, these rare
events do not contribute much to $\langle 1 - e^{-F} \rangle$, since
this expression saturates for large $F$.  Due to these large
fluctuations it is therefore possible that one can observe events in
which there are large saturation effects, even though the average
scattering is still weak. Although such rare events are less important
for the total cross section, they are very essential for the
diffractive cross section.

In leading order the dipole splitting in \eqref{eq:dipkernel1}
diverges for small dipoles, and therefore the number of small dipoles
depends strongly on the necessary cutoff. As pointed out in
\cite{Salam:1995uy} a very essential source of the large fluctuations
is also the occasional creation of a very large dipole. Such a large
dipole has a large probability to split, and the most likely scenario
is that it splits into one very small dipole and one dipole which is
almost equally large. The process is then iterated and the result is a
``jet'' of many small dipoles.


We note, however, that the very large fluctuations observed by Mueller
and Salam are strongly reduced by non-leading effects.  As
demonstrated in \cite{Avsar:2005iz} energy--momentum conservation has a
very strong influence. The production of small dipoles is suppressed
by the conservation of the positive light-cone momentum $p_+$, while
the large dipoles are suppressed by the conservation of $p_-$.
Another non-leading effect comes from the running coupling \as.
As discussed in section \ref{sec:running}
the relevant scale in the dipole splitting is
determined by min$(r, r_1, r_2)$ where $r$ is the mother dipole which
splits into $r_1$ and $r_2$. This suppresses not only the production
of very small dipoles, but also fluctuations where a small dipole
splits producing two very large dipoles. We thus conclude that both
effects contribute to a suppression of very small or very large
dipoles, and therefore also of the abovementioned ``jets'' radiated
from an occasional large dipole.  As a result we find in our
calculations that $\langle F^k \rangle$ grows like $k!$ rather than
$(k!)^2$. The ratios $\langle F^k\rangle /(k\cdot\langle
F^{k-1}\rangle)$ for dipole-dipole scattering at $Y=10$ are
approximately equal to 1.2 for all $k$ between 5 and 9 (larger
$k$-values need very high statistics, and are therefore difficult to
simulate). However, although the fluctuations in the cascade evolution
are strongly tamed by non-leading effects, they are still very
important, and have a large effect on the diffractive cross sections,
as seen in section \ref{sec:sources} and in the results presented in
the next section.


\section{Results on diffraction and elastic scattering}
\label{sec:res}


\subsection{Diffraction in $pp$ collisions}
\label{sec:ppres}

\FIGURE[t]{
  \hspace{-0.07\linewidth}\epsfig{file=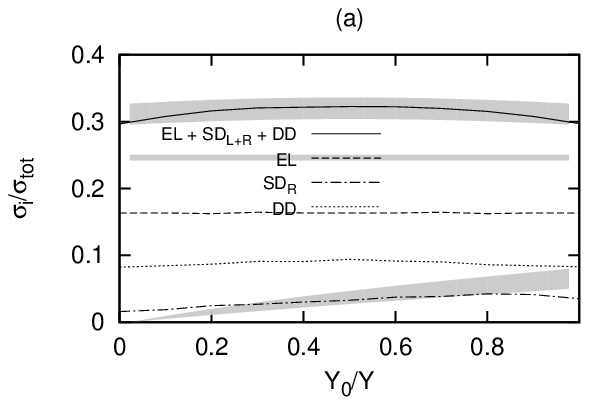,width=0.6\linewidth}\hspace{-0.15\linewidth}\epsfig{file=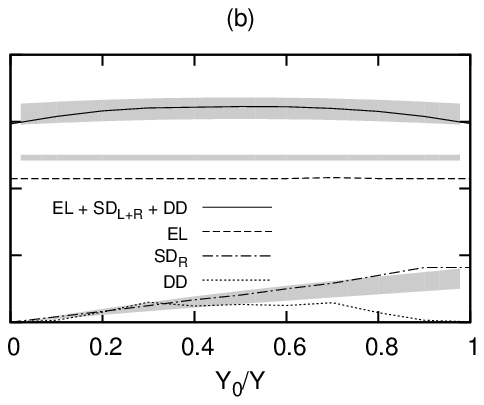,width=0.6\linewidth}\hspace{-0.07\linewidth}
  \caption{\label{fig:ppdiffract} The ratio between the total diffractive
    and the total cross section (solid line) together
    with the contribution from elastic (dashed), single-right
    (dash-dotted) and double diffractive (dotted) cross sections at 1.8 TeV,
    obtained including (a) and excluding (b) fluctuations in the
    initial proton wavefunction. In both cases the lower error band is an
    estimate from CDF data on single diffraction \cite{Abe:1993wu},
    the middle band is the CDF elastic cross section \cite{Abe:1993xy}
    and the upper is a sum of the two contributions. (Thus the contribution
    from double diffraction is not included in the CDF result.) }}

In figure \ref{fig:ppdiffract} we show the ratio of the total
diffractive, single diffractive and elastic cross sections to the
total cross section at 1.8~TeV as a function of the Lorentz frame
used.  Figure \ref{fig:ppdiffract}a shows the result obtained with the
initial proton wavefunction in \eqref{eq:pwave1}, while
figure \ref{fig:ppdiffract}b shows the upper limit obtained by
integrating over the initial proton wavefunctions after taking the square in
\eqsref{eq:sigmael}-(\ref{eq:sigmasdl}). In these figures 
we also show results from the CDF collaboration (not including double
diffraction).

The results in figure \ref{fig:ppdiffract}a do not agree well with the
data. The elastic cross section is too low. The single diffractive
cross section $\sigma_{SD}^R$
does not go to zero when $Y_0 \rightarrow 0$. In this limit $M_X^2 
\rightarrow m_p^2$ and there should be no phase space for diffractive
excitation. The double diffractive excitation is about 9\%, which
is unrealistically high, and this cross section should also approach
zero in the limits $Y_0 \rightarrow 0$ or $Y_0 \rightarrow Y$. All these
features illustrate the problems with our initial proton wave 
function, discussed above. This formalism presumes that there
are other hadronic states,  
which have wavefunctions orthogonal to the proton wavefunction,
and which have approximately the same mass.
The limiting results in fig~\ref{fig:ppdiffract}b, obtained when the 
fluctuations in the initial proton wave function are neglected,
do not have this problem. We also see that these results agree quite
well with the experimental estimates, supporting the assumption that
the initial wavefunction fluctuations have a small effect compared to the
fluctuations in the cascade evolution and the impact parameter dependence.

For the elastic cross section we also note (as was claimed above) that
it is approximately frame-independent, as it should be. The upper
limit shown in figure \ref{fig:ppdiffract}b is around 22\%, which
agrees well with the value ($22.02\pm 0.78$)\% from E811
\cite{Avila:1998ej}, while the value from CDF, ($24.6 \pm 0.4$)\%
\cite{Abe:1993xy} is a bit larger.

The single diffractive cross
section $\sigma_{SD}^R$ ($\sigma_{SD}^L$) is increasing (decreasing)
when $Y_0$ is increased, and in the model we make the identification of the
diffractive masses:
\begin{equation}
M_X^{2(R)} = e^{Y_0}\,\, \mathrm{GeV}^2, \quad
M_X^{2(L)} = e^{Y-Y_0} \,\,\mathrm{GeV}^2,
\label{eq:MX}
\end{equation}
where for 1.8 TeV we have $Y=\ln(s/1\mathrm{GeV}^2) \approx 15$.
In figure \ref{fig:ppdiffract} we also show results obtained from the CDF 
parameterization of single diffractive excitation \cite{Abe:1993wu}:
\begin{equation}
\frac{d \sigma_{SD}^R}{d \ln M_X^{2(R)}} = 
\frac{1}{2}\frac{D}{(b_0+0.5\ln(s/M_X^{2(R)}))}\biggl (\frac{s}{M_X^{2(R)}}\biggr )^\epsilon
\label{eq:param}
\end{equation} 
with $D=2.54\pm0.43$~mb, $b_0=4.2\pm0.5$, and $\epsilon=0.103\pm0.017$.
Similar results are also presented by the E710 collaboration 
\cite{Amos:1992jw}, although with somewhat larger errors. Our results in 
figure \ref{fig:ppdiffract}b agree quite well with the data, even if they are a
little high for the largest excited masses. Besides not going to
0 when $Y_0\rightarrow 0$, the result in figure \ref{fig:ppdiffract}a also
has a much too slow variation with $Y_0$, meaning a too low value for
$d \,\sigma_{SD} /d\,\mathrm{ln} M_X^2$.

For double diffraction our result for $Y_0=7.5$, which corresponds to
a central gap, is 2.0 mb. Experimental data exist for 900 GeV from the
UA5 collaboration at the CERN Sp\=pS collider \cite{Ansorge:1986xq}. 
Our result at this energy is 1.8 mb, which is consistent with the 
experimental result $4.0\pm2.2$ mb. Our results can also be compared to
the model of Goulianos \cite{Goulianos:1995wy}, who argues 
that $\sigma_{DD}$ should decrease with energy, due to saturation effects, 
from around 1.6~mb at 900~GeV to around 1.3~mb at 1.8~TeV.
We also note that our
result is consistent with a factorized dependence on the two masses,
as expected from Regge formalism:
\begin{equation}
\frac{d \sigma_{DD}}{d M_X^{2(R)} \, d M_X^{2(L)}} = Const. \cdot
f( M_X^{2(R)})\cdot f( M_X^{2(L)}),
\end{equation}
where $f(M_X^2)$ denotes the distribution for single diffraction in 
\eqref{eq:param}. (Note that double diffraction is not included in the 
CDF data in figure \ref{fig:ppdiffract}.)

As seen in figure \ref{fig:pptotrmaxlqcd} the same total cross 
section can be obtained with different sets of values for 
$\Lambda_{QCD}$ and $r_{max}$. In figure \ref{fig:ppelrmaxlqcd} we see that
varying these parameters, keeping
$\sigma_{tot}$ constant, does modify the elastic cross section somewhat.
However, our upper limit is still close to the data, leaving little
room for a contribution from the initial proton wave function.
\FIGURE[t]{
  \epsfig{file=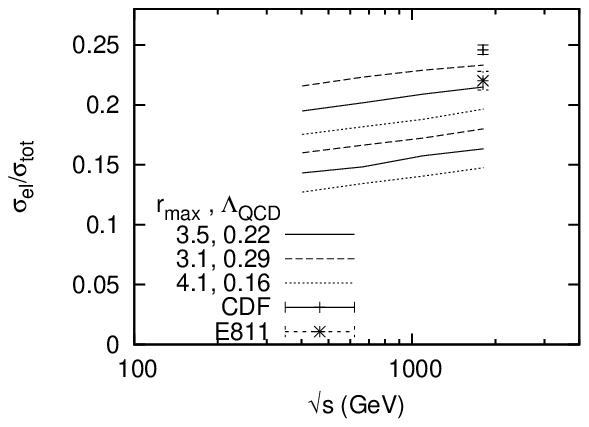,width=0.6\linewidth}
  \caption{\label{fig:ppelrmaxlqcd} The ratio of the elastic to the
    total cross section in $pp$ collisions as a function of $\sqrt{s}$
    for various values of $r_{max}$(GeV$^{-1}$) and
    $\Lambda_{QCD}$(GeV). Lower curves are obtained including
    fluctuations in the initial proton wavefunction, while the upper
    curves excludes these fluctuations. The data points are from CDF
    \cite{Abe:1993xy} and E811 \cite{Avila:1998ej}.}
}

\FIGURE[t]{
  \epsfig{file=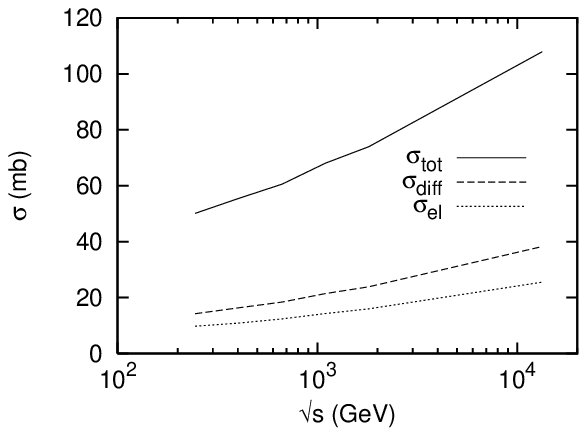,width=0.6\linewidth}
  \caption{\label{fig:ppdiffract2} The total (full line), total
    diffractive (including elastic) (dashed line) and elastic cross 
    sections (dotted line)
    as a function of collision energy,$\sqrt{s}$, in $pp$ collisions. Here the
    diffractive cross section is evaluated at $Y_0/Y =
    0.5$ \ie\ with a central gap in the cms.  } }

Finally we show in figure \ref{fig:ppdiffract2} the energy dependence of
the total, the total diffractive (including elastic scattering),
and the elastic cross sections, including our predictions for the LHC. 
(The values presented correspond to those in fig~\ref{fig:ppdiffract}b,
\ie\ to the more realistic results obtained neglecting the fluctuations in 
the initial proton wave function.) The diffractive cross section is calculated
in the cms with $Y_0=Y/2$, which demands a central gap
and implies that the diffractive excitations
are limited by $M_X^2 < \sqrt{s}\cdot 1\mathrm{GeV}$. The values we predict
for the LHC are 108, 38, and 26~mb respectively.

\subsection{Diffraction at HERA}
\label{sec:herares}

Diffractive excitation has been measured at HERA by the ZEUS
\cite{Chekanov:2005vv} and H1 \cite{Adloff:1997sc} collaborations with two
different methods. One is based on an observed rapidity gap. The ZEUS
data obtained with this method \cite{Chekanov:2005vv} give the cross
section integrated over all diffractively excited protons with mass
$M_X^{(p)} < 2.3$ GeV. Assuming that the contribution from events
where the proton is excited beyond this limit is small, and can be
neglected, the result of this method for $M_X^{(\gamma)} <
M_{X,max}^{(\gamma)}$ corresponds to our model calculations for
$\sigma_{diff}$ at $Y_0=\ln(M_{X,max}^{2(\gamma)})$:
\begin{eqnarray}
  \sigma_{diff}(M_{X,max}^{2(\gamma)}) =
  \int^{\mathrm{ln}M_{X,max}^{2(\gamma)}} d\,\mathrm{ln}M_X^{2(\gamma)} 
  \frac{d\sigma_{diff}}{d\,\mathrm{ln}M_X^{2(\gamma)}} =
  \sigma_{diff}^{(model)} (Y_0=\ln(M_{X,max}^{2(\gamma)})).
\end{eqnarray}
The results are shown in figure \ref{fig:diffres} and we see a very
good agreement with data, although there is a tendency for our cross
sections to decrease a bit too slowly with $Q^2$.

\FIGURE[t]{
  \epsfig{file=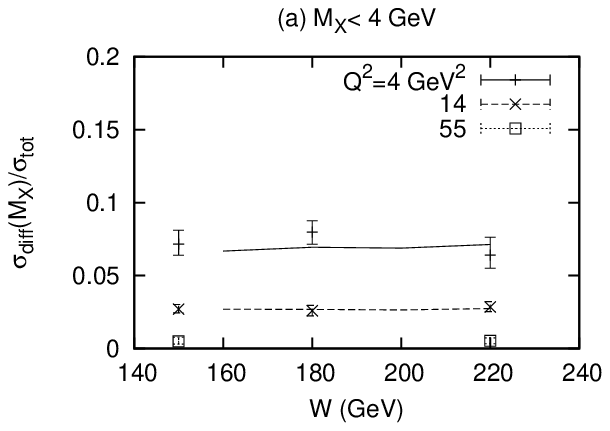,width=0.5\linewidth}%
  \epsfig{file=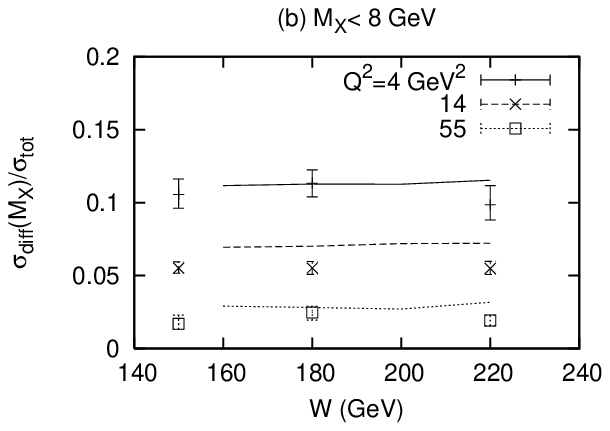,width=0.5\linewidth}
  \epsfig{file=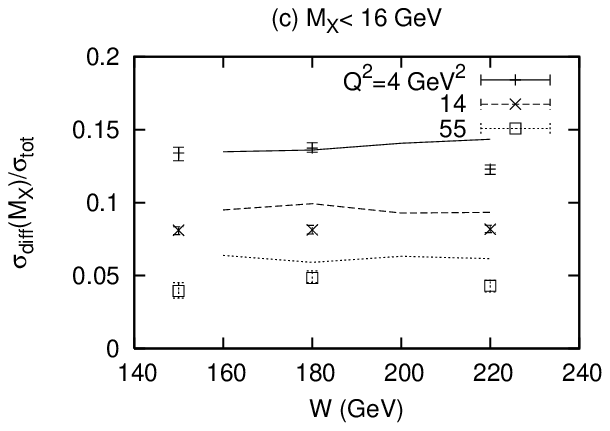,width=0.5\linewidth}%
  \epsfig{file=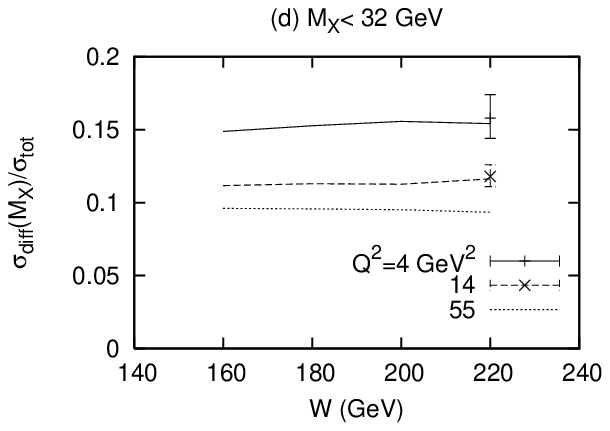,width=0.5\linewidth}
  \caption{\label{fig:diffres} The ratio of the total diffractive
    cross section to the total cross section as a function of $W$, for
    $M_X<4$ (a), $8$ (b), $16$ (c) and $32$~GeV (d). Our results are
    compared to ZEUS data \cite{Chekanov:2005vv} for $Q^2=4$ (full
    lines and + points), $14$ (dashed lines and x points) and
    $55$~GeV$^2$ (dotted lines and open squares). } }

The cross section for single diffractive excitation of the photon can
also be calculated in our model, and in figure \ref{fig:sdres} we
present the ratio wrt.\ the total diffractive cross section as
a function of $W$ for different $Q^2$ and $M_X$. In
\cite{Chekanov:2005vv} the ZEUS collaboration estimated this ratio to
be $0.70\pm0.03$, by comparing a parameterization\footnote{Using a
  modified version of the model in \cite{Bartels:1998ea}.} of their
diffractive data to results from their leading proton spectrometer.
This result is obtained using the assumption that the ratio is
independent of $W$, $Q^2$ and $M_X$.  Comparing with figure
\ref{fig:sdres} we find that our result is consistent with the ZEUS
number, but that we predict that the ratio actually does have a small
dependence on $M_X$ and $Q^2$.

\FIGURE[t]{
  \epsfig{file=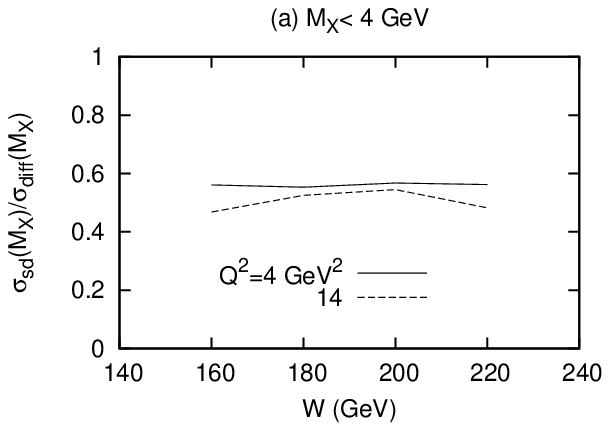,width=0.5\linewidth}%
  \epsfig{file=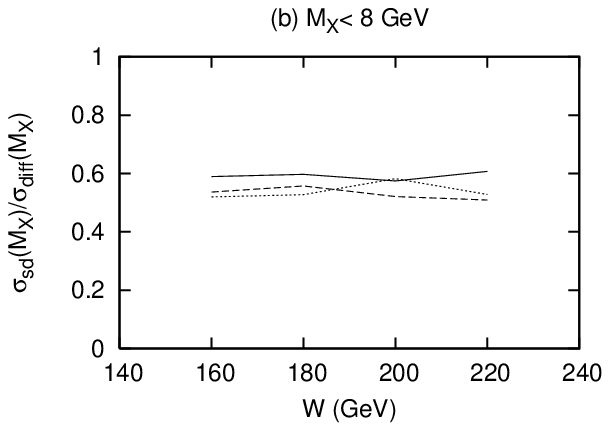,width=0.5\linewidth}
  \epsfig{file=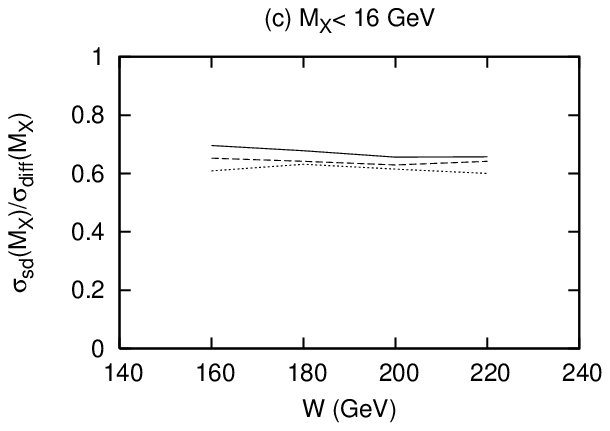,width=0.5\linewidth}%
  \epsfig{file=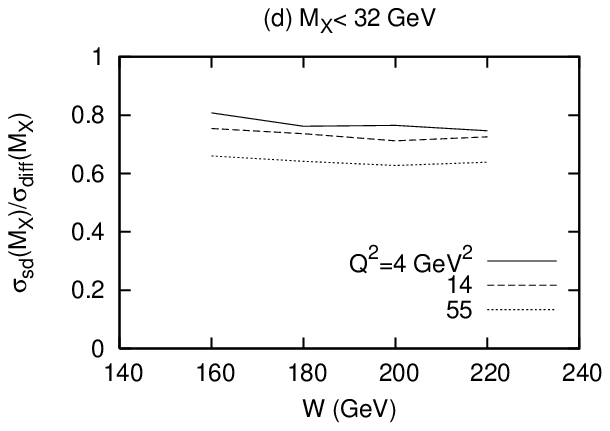,width=0.5\linewidth}
  \caption{\label{fig:sdres} The ratio of the single diffractive cross
    section to the total diffractive cross section as a function of
    $W$, for $M_X<4$ (a), $8$ (b), $16$ (c) and $32$~GeV (d) for
    $Q^2=4$ (full lines), $14$ (dashed lines) and $55$~GeV$^2$ (dotted
    lines).  } }

\section{Conclusions and Outlook}
\label{sec:conc}

We have in earlier papers presented an extension of Mueller's dipole
cascade model, which includes subleading effects from energy
conservation and running coupling as well as colour suppressed effects
from pomeron loops. The model is also implemented in a MC simulation
program, which simplifies the comparison between theoretical ideas
and experimental data, and allows more detailed studies of
important non-leading effects. Calculations of total cross sections
agree very well with experimental results for $pp$ collisions and deep 
inelastic electron scattering.

To gain further insight into small-$x$ evolution and saturation,
we have in this paper first presented a number of improvements of 
the model, in particular related to the confinement mechanism, and
thereafter applied the model to elastic scattering and diffractive
excitation, where we specifically study the effects of different
sources for fluctuations.

A consistent treatment of confinement effects is achieved by replacing the
infinite range Coulomb potential in the dipole splitting and in the
dipole--dipole scattering with a screened Yukawa potential. By
equating the screening length, $r_{\max}$, with the size of the proton
entering into its wavefunction, we were able to get a good,
boost-invariant description of the $pp$ and $\gamma^\star p$ total
cross sections for a wide range of energies, using basically only two
parameters, $r_{\max}$ and $\Lambda\sub{QCD}$. This new treatment of
confinement has effects on the boost invariance of the model,
further improving the earlier, almost frame independent, results.

Our treatment of diffraction is based on the formalism of Good--Walker 
and Miettinen--Pumplin. The cross sections for
elastic scattering and diffractive excitation are here determined by
the fluctuations in the interaction probability between different events.
Contrary to other calculations, we can in our model easily
consider all different sources for such fluctuations; those stemming from
the dipole cascade evolution, the dipole--dipole scattering, from the
impact parameter dependence, and from the initial photon
and proton wavefunctions. We find that all of these sources give
important contributions, apart from the initial proton wavefunction, and
together they give a very good description of data on elastic and
single- and doubly-diffractive scattering in both $\gamma^\star p$ and
$pp$ collisions. We must, however, admit that we do not have a realistic 
quantum-mechanical description of the proton state in terms
of dynamical variables. Here data are best reproduced if the contribution
from the fluctuations in the initial proton state are small compared 
to the other contributions.

In a future publication we will use our model to study the quasi-elastic 
reactions $\gamma^\star p\to Vp$ and deeply inelastic Compton
scattering (DVCS). In the future we also want to develop
the model further to be able to describe exclusive multi-particle
final states. This needs, however, a recipe for how to handle the virtual 
dipoles, those which do not participate in the collision and therefore
cannot come on shell and give final state hadrons. To describe
particle production in the proton fragmentation regions would also
need a much improved description of the initial proton state.

\bibliographystyle{utcaps}
\bibliography{/home/shakespeare/people/leif/personal/lib/tex/bib/references,refs}

\end{document}